\documentclass[aps,prb,reprint,groupedaddress]{revtex4-1}

\usepackage{graphics}

\begin{document}


\title{Finite-size Kosterlitz-Thouless transition in 2DXY Fe/W(001) ultrathin films}

\author{J. Atchison}
\author{A. Bhullar}
\author{B. Norman}
\author{D. Venus}
\email{[corresponding author] venus@physics.mcmaster.ca}
\affiliation{Department of Physics and Astronomy, McMaster University, Hamilton, Ontario, Canada}

\date{\today}

\begin{abstract}
Magnetic susceptibility measurements of 3-4 ML Fe/W(001) ferromagnetic films demonstrate that this is a 2DXY system in which a finite-size Kosterlitz-Thouless (KT) transition occurs.  The films are grown in ultrahigh vacuum and their magnetic response is measured using the magneto-optic Kerr effect (MOKE).  The analysis of many independently grown films shows that the paramagnetic tail of the susceptibility is described by $\chi(T)=\chi_0 \exp \bigr{(}B/(T/T_{KT}-1)^a\bigl{)}$, where $a=0.50\pm0.03$ and $B=3.48\pm0.16$, in quantitative agreement with KT theory.   Below the finite-size transition temperature $T_C(L)$, the behaviour is complicated by dissipation (likely related to the re-emergence of four-fold anisotropy and magnetic domains).  A subset of measurements with very small dissipation most closely represents the idealized system treated by theory.  For these measurements, there is a  temperature interval of order tens of K between the fitted Kosterlitz-Thouless transition temperature and the finite-size transition temperature, in agreement with theory.  The normalized interval $T_C(L)/T_{KT} -1=0.065\pm0.016$ yields an estimate of the finite size $L$ affecting the film of order micrometers.  This gives experimental support to the idea that even a mesoscopic limitation of the vortex-antivortex gas results in a substantial finite-size effect at the KT transition.  In contrast, fitting the paramagnetic tail to a power law, appropriate to a second order critical transition, gives unphysical parameters.  The effective critical exponent $\gamma_{eff} \approx 3.7 \pm 0.7$ does not correspond to a known universality class, and the fitted transition temperature, $T_{\gamma}$, is much further below the peak in the susceptibility than is physically reasonable. 
\end{abstract}


\maketitle

\section{Introduction}
The concept of a topological phase transition occurring in two dimensional (2D) systems was put forward in a series of articles by Berezinskii,\cite{Berezinskii} Kosterlitz and Thouless\cite{Kosterlitz1,Kosterlitz2} more than 45 years ago. These ideas have assumed an important  role in condensed matter physics.\cite{Jose}  The original paper\cite{Kosterlitz1} considered topological transitions in the melting of a 2D lattice, in neutral superfluids, and in a 2DXY ferromagnetic film, with a quantitative analysis using the 2DXY ferromagnet as a model system following soon thereafter.\cite{Kosterlitz2}  Subsequent experimental investigations of the Kosterlitz-Thouless (KT) transition have concentrated on superfluids and superconducting Josephson junction arrays,\cite{Goldman} with little work on 2DXY ferromagnetic films.  Because each system offers a different window through which to view the KT transition, there is a strong motivation to confirm experimentally that a 2D ultrathin film ferromagnet exhibits the transition.  A more detailed understanding of the implications of KT theory in non-ideal, physical realizations of the 2DXY model can then be gained through the experimental study of these accessible, easily prepared systems.

Earlier experimental work on 2DXY spin systems studied planar three dimensional (3D) antiferromagnets and ferromagnets, where the exchange coupling within a 2D plane is much larger than that between planes.\cite{Bramwell1,Taroni} These samples are suitable for neutron scattering techniques.   A series of theoretical articles by Bramwell, Holdsworth and co-workers\cite{Bramwell1,Bramwell2,Archambault2} pointed out the essential role of finite-size effects in the KT transition in these systems, and showed that this results in a temperature range where the magnetization  $M(T)$ (or the staggered magnetization in antiferromagnets) scales  like a power law with an effective exponent $\beta_{eff}$.  This range ends at a finite-size transition temperature $T_C(L) > T_{KT}$, where $L$ is the finite size affecting the transition.  This power law scaling has been observed in a number of compounds,\cite{Bramwell1,Taroni} and a detailed neutron scattering study of the antiferromagnet Rb$_2$CrCl$_4$ showed in addition an internally consistent analysis of the correlation length and susceptibility in terms of a finite-size KT transition.\cite{Nielson}

The connection between these ideas and truly 2D ultrathin ferromagnetic films, as envisioned in the original papers by Kosterlitz and Thouless, has been made by compiling published experimental determinations of $\beta_{eff}$ for ultrathin epitaxial metal films on substrates of different symmetries.\cite{Taroni}  For metallic films grown on the faces of cubic substrates, the distribution of exponent values is bimodal for (001) and (111) substrates, with clusters near the Ising value $\beta$ = 0.125 and the 2DXY effective value $\beta_{eff}$=0.231.  For (110) substrates, the measurements cluster near the Ising exponent .  This is important, but somewhat ambiguous, evidence of KT behaviour of 2DXY ferromagnetic films.  The authors' primary point is that the four-fold anisotropy of the ferromagnetic films grown on (001) substrates is not strong enough to move them into a completely new universality class where $\beta$ is much larger.

The present article investigates the magnetic susceptibility of ultrathin Fe/W(001) films, and compares it to the theory of the finite-size KT transition.  There are a number of advantages to measuring the susceptibility, as compared to the magnetization.  Whereas the magnetization signal disappears as the transition is approached, the susceptiblity signal exhibits a peak that can be studied in detail.  Thus, in addition to the value of exponents predicted by the theory, the peak shape in the paramagnetic phase can be compared to the distinct exponential temperature dependence of KT theory.   Since a applied static field is not used to induce a single domain state, the susceptibility in the temperature range between $T_{KT}$ and $T_C(L)$ can also be used to probe the role of domain walls and vortex-antivortex pairs in mediating the emergence of isotropy from the underlying, low temperature, anisotropic state.  
 
To our knowledge, there is one published measurement of the susceptibility of a candidate ferromagnetic ultrathin film 2DXY system, and this is also for Fe/W(001).\cite{Elmers1}  This measurement was made as the difference of magnetization curves in which a slightly different d.c. magnetic field is applied.  The high temperature tail of the curve was compared to the predictions of finite-size KT theory, but the result is inconclusive.  The current article reports on the growth, measurement and quantitative analysis of the magnetic susceptibility of many independent Fe/W(001) films in the paramagnetic phase.  The excellent quantitative agreement with finite-size KT theory, and incompatibility with the predictions of a second-order critical transition, shows that this is a 2DXY system that exhibits a KT transition.  This accessible ultrathin film system offers new opportunities to study topological phases and transitions by straightforward measurements of the magnetic susceptibility.


%

\section{Theoretical description}
The Mermin-Wagner theorem\cite{Mermin} proves that a 2D array of in-plane spins with nearest neighbour exchange coupling $J$, and no anisotropy, cannot order at finite temperature in the thermodynamic limit.  If only spin wave excitations are considered, the susceptibility always diverges and the magnetization is zero. However, because the 2DXY model admits another spin excitation, the vortex, it can undergo a type of transition mediated by the unbinding of vortices and antivortices.\cite{Kosterlitz2}

Below the transition temperature $T_{KT}$, the vortices and antivortices form bound pairs that represent only a small proportion of the spin system, such that the spin waves are dominant.  With increasing temperature the pairs are bound more loosely and begin to create an effective medium with reduced exchange coupling $J_{eff}$.  As the thermal energy approaches the binding energy of the pairs, there is a highly non-linear feedback, where reducing the exchange coupling produces more loosely bound pairs, which reduces the coupling, and so on.  The exchange coupling is driven to $J_{eff}/ kT = \pi/2$ at $T_{KT}$, and then directly to zero above the transition as the vortex-antivortex pairs unbind and proliferate.

Above the KT transition, the susceptibility becomes finite and varies as\cite{Kosterlitz2}
\begin{equation}
\label{chi}
\chi(T) \sim \xi^{2-\eta},
\end{equation}
where the exponent $\eta = 1/4$ at the transition temperature, and the correlation length $\xi$ does not diverge like a power law, but as
\begin{equation}
\label{xi}
\xi \sim \exp \Bigl{(}\frac{b}{\sqrt{\frac{T}{T_{KTB}}-1}}\Bigr{)}.
\end{equation}
The best estimates\cite{Nielson,Gupta} of $b$ are 1.8 to 1.9.

In the spin wave region of real systems below $T_{KT}$, the magnetization falls so slowly with the system size that the thermodynamic limit is not reached even for macroscopic samples.\cite{Bramwell1,Bramwell2}  For this reason, the magnetization of a 2DXY ferromagnetic film exhibits a substantial magnitude below $T_{KT}$ due to finite-size effects.  This can then be aligned by a four-fold anisotropy to create a net magnetization.   Furthermore, the distribution of vortex-antivortex pairs is truncated by the finite size, $L$, of the system, so that $J_{eff}/kT$ is renormalized more slowly than in an infinite system.   It reaches the value of $\pi/2$ at temperature $T^*$, where
\begin{equation}
\label{T*}
\frac{T^*-T_{KT}}{T_{KT}}=\frac{b^2}{4(\ln L)^2}.
\end{equation}
$T_{KT}$ continues to denote the transition temperature for the infinite system.  Above this temperature, $J_{eff}$ falls more gradually than in the infinite system, such that there is a finite-size transition\cite{Bramwell1} at $T_C(L)$. Setting $\xi(T) \sim L$ in eq.(\ref{xi}) yields:
\begin{equation}
\label{TC}
\frac{T_C(L)-T_{KT}}{T_{KT}}=\frac{b^2}{(\ln L)^2}.
\end{equation}

Between $T^*$ and $T_C(L)$, the magnetization falls to zero like a power law\cite{Bramwell1} with an effective exponent $\beta_{eff}$, so that
\begin{equation}
\label{beta}
M(T) \sim \Bigl{(}1-\frac{T}{T_C(L)}\Bigr{)}^{\beta_{eff}}.
\end{equation}
The fact that this behaviour has been observed in a number of 2D ferromagnetic films\cite{Taroni} suggests that although the four-fold anisotropy is essential in creating a net magnetization, it does not determine the critical properties leading up to the transition.  In this same region, the calculated susceptibility\cite{Archambault1} increases with temperature due to softening of the spin waves, until it reaches a maximum at $T_C(L)$.  Above this temperature, the distribution of unbound vortices creates a paramagnetic state where the susceptibility once again varies as in eq.(\ref{chi}), with the correlation length in eq.(\ref{xi}) determined by $T_{KT}$ for the infinite system.

The paramagnetic tail of the experimental susceptibility above $T_C(L)$ is therefore expected to vary as
\begin{equation}
\label{chiexp}
\chi(T) = \chi_0 \exp \Bigl{(}\frac{B}{(\frac{T}{T_{KT}}-1)^a}\Bigr{)},
\end{equation}
where $a=1/2$ and $B=(2-\eta)b$.    While $\eta$=1/4 at $T_{KT}$, it may be as low as zero in the paramagnetic state.\cite{Nielson}  Thus $B$ may range from about 3.2 to 3.8 .  In the paramagnetic region, the gas of free vortices should respond to an applied field without dissipation, so that the imaginary component of the susceptibility will be very small or absent.  Eq.(\ref{chiexp}) is to be compared with the corresponding quantity in a second order critical transition,\cite{Stanley}
\begin{equation}
\label{power}
\chi(T)=\chi_0 \Bigl{(}\frac{T}{T_{\gamma}}-1 \Bigr{)}^{-\gamma},
\end{equation}
where $\gamma$ is the critical exponent and the second-order transition temperature is $T_{\gamma}$.  For a 2D Ising transition, $\gamma = 7/4$; for a four state Potts model\cite{Entig}, $\gamma = 7/6$.

The form of the measured susceptibility below $T_C(L)$ is not as clear.  In Ising thin film ferromagnets, where  two-fold anisotropy is present both above and below the second-order transition, a number of experiments have explored the dissipative contribution to the susceptibility using techniques where an applied static field is not used to induce a single domain state.  In both thicker bulk-like films\cite{Bovensiepen} and 2D films\cite{Dunlavy,Dunlavy2} the anisotropy is expressed immediately upon passing through the transition from above, by the formation of magnetic domains.  Dissipation due to domain wall motion modifies the low temperature side of the narrow susceptibility peak, and creates a similarly narrow accompanying peak in the imaginary component of the susceptibility.  The temperature of the peak of Im$\chi(T)$ typically occurs within 1 or 2 K of the maximum of Re$\chi(T)$.

In contrast, in the n-fold infinite KT transition, the underlying anisotropy is relevant below the transition and there is emergent isotropy above it.\cite{Ortiz}  The process by which emergent isotropy is destroyed as the temperature is reduced, may not be closely analogous to the process by which domains develop in the presence of a continuous anisotropy in an Ising transition.  This is particularly the case for the four-fold finite-size KT transition, where the transition is ``smeared out" between the temperatures $T^*$ and $T_C(L)$.  At $T_C(L)$ the magnetic excitations are free vortices, and some tens of K lower at $T^*$ the magnetic excitations are domain walls.  The manner in which the system crosses over from one regime to the other has not been explored;  a reasonable conjecture is that Im$\chi(T)$ is dominated by domain wall motion in a lower temperature range, and by vortex-antivortex unbinding in a higher temperature range.  Simulations and theoretical calculations of the susceptibility have not included dissipative contributions or the effect of domain walls, although there is some work on non-equilibrium quenching\cite{Berthier} well below $T_{KT}$ and non-equilibrium relaxation\cite{Ozeki} near $T_{KT}$.

The experiments described here concentrate on the high temperature, paramagnetic state  where detailed theoretical models make explicit predictions.  The detailed characterization of the temperature range below $T_C(L)$ will be the focus of future experiments.

\section{Experimental methods}

The growth of Fe/W(001) films has been studied by a number of groups.\cite{Berlowitz,Zhou,Jones,Elmers2,Niu}  After some initial uncertainty, it has been established that 2ML of Fe on W(001) are thermally stable up to 700 K, whereas thicker films are stable to lower temperatures.  This permits a straightforward thickness calibration of films grown by evaporation in UHV using the Auger electron spectroscopy (AES) signal from the W substrate during deposition.  The amplitude of the W AES peaks in the energy range 150 to 190 eV shows a clear break between two linear regions as a function of deposition time, if the films are annealed to 700 K.\cite{Jones}  This thickness is interpreted as the completion of 2 ML of Fe\cite{Elmers2}, giving a calibration to within 5\%.  A recent magnetic microscopy study\cite{Niu} confirmed layer growth for 3 to 4 layers, after which the formation of three dimensional islands begins.  The current study uses  films of  3 to 4 ML Fe, because the same microscopy study showed that, while the magnetic domain structure is sensitive to thickness and preparation, large magnetic domains oriented along the principle axes were observed in films in this thickness range.  The films in the present article were grown in two stages -- the first 2 ML was grown at room temperature and annealed to 600 K to promote wetting and smoothing, and then an additional 1 or 2 ML was grown at room temperature.  The entire film was then annealed to 460 K to ensure stability during $\chi(T)$ measurements that can extend as high as 450 K.  The temperatures throughout the experiments were measured using a W-Rh thermocouple embedded in the edge of the W substrate crystal.

\begin{figure}
\scalebox{.45}{\includegraphics{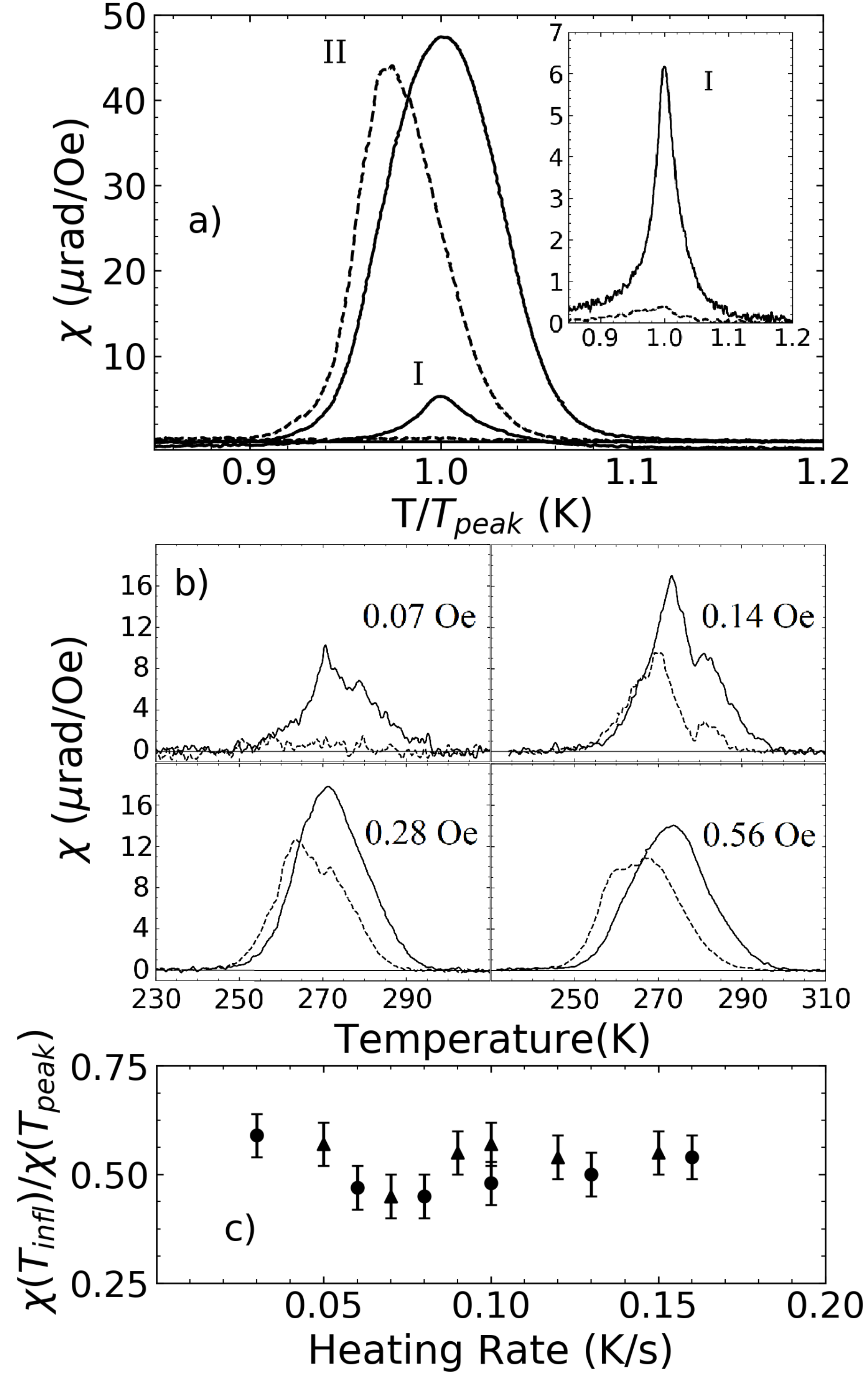}}
\caption{\label{fig1}a) Regularly shaped magnetic susceptibility measurements fall into two categories, each occuring in roughly one third of the films.  Type I signals (also shown in the inset) have a small magnitude and a very small imaginary component.  Type II signals are an order of magnitude larger and have a large imaginary component.  The measurements were made with an a.c. field amplitude of 0.56 Oe.  b) The magnetic susceptibility of a type II film is shown for selected values of the a.c. field amplitude. c) The effect of the heating rate on the high temperature tail of $\chi(T)$ is assessed by how far down the peak the point of inflection, at $T_{infl}$, occurs.  The findings for two different films are indicated by the circular and triangular symbols.}
\end{figure}

The magnetic susceptibility was measured using the magneto-optic Kerr effect (MOKE) in the longitudinal geometry.\cite{Qiu}  Details of the apparatus and procedures are described in ref.(\onlinecite{Arnold}) and (\onlinecite{Arnold2}).  HeNe laser light passes through a polarizing crystal, enters the chamber through a UHV window, scatters from the sample at $45^o$, exits through a second UHV window and a second polarizing crystal that is almost crossed with the first.  The transmitted light then falls on a photodiode.  An optical  compensation technique is used to ensure that the light falling on the second polarizer is linearly polarized, so that the sensitivity of the method is optimized.   A pair of wire coils attached to the sample holder is used to create a small in-plane a.c. field within the scattering plane.  The output of the photodiode is connected to a lock-in amplifier that detects the Kerr rotation in phase (Re$\chi$) and out of phase (Im$\chi$) with the a.c. magnetic field.  The sample can be rotated about its normal, so that any in-plane component of the magnetization can be probed.  In the present experiments, the field was aligned with either the [100] or [010] axis.

The magnetic susceptibility measurements of the Fe/W(001) films fall into three qualitative groups.  Two of these are shown in fig.(\ref{fig1}a), using a solid line for Re$\chi(T)$ and a dashed line for Im$\chi(T)$.  Roughly 1/3 of the measurements are like the curve labelled I, and shown in the inset to fig.(\ref{fig1}a).  These have a relatively small magnitude and a very small imaginary component.  These will be referred to as type I signals.  In another 1/3 of the measurements, the peak in Re$\chi(T)$ is about an order of magnitude stronger, the peak is wider, and it is accompanied by strong peak in Im$\chi(T)$.  These are termed type II signals.  Finally, the remaining 1/3 of the measurements (not shown) have a complex form below the peak temperature that is different from film to film.   The shape of the susceptibility peak, as indicated by these categories, is not correlated to the film thickness, or growth protocol in the small range of thicknesses that has been investigated here.   All the measurements are qualitatively different than the susceptibility of the 2D Ising system Fe/W(110).\cite{Back,Elmers3,Dunlavy,Dunlavy2}  Even the narrower peaks in type I signals have a normalized full width at half maximum $\Delta T/T_{peak} \approx 0.050$, which is more than twice the value of 0.018 observed in the second order critical transitions of Fe/W(110).  This is despite the fact that the susceptibility of the latter has a substantial imaginary component. 

In order to determine an appropriate a.c. field amplitude for the measurements, the susceptibility of a type II film was measured for a range of amplitudes.  A selection of these susceptibility curves is shown in fig.(\ref{fig1}b).  When the a.c. field is reduced to 0.07 Oe, the imaginary component of the susceptibility is greatly reduced, and becomes of similar size to that observed in type 1 measurements made with large fields (0.56 Oe in the inset to fig.(\ref{fig1})a).  While the signal is very noisy, there is an indication of internal structure, with Re$\chi(T)$ showing two maxima.  This is confirmed when the a.c. field magnitude is in the range 0.10 to 0.21 Oe and the noise is reduced, with both Re$\chi$ and Im$\chi$ showing a pronounced double peak structure.   For a field amplitude of 0.28 Oe, the double peak structure is no longer resolved, with only a slight indication in Im$\chi$.  When the field is increased to 0.56 Oe, there is a single, broad peak, as in part a) of the figure where this field value is used.

We speculate that the double peak structure of the susceptibility observed for small a.c. field amplitude is related to the crossover from a system dominated by domain walls  to one dominated by vortex-antivortex pairs over a temperature range of about 20 K.  If this is the case, then the type I measurements would represent films where the smaller, narrower peak at higher temperature is present, but the larger peak at lower temperature is mostly absent.  In this interpretation, this might be due to films with different types of domain structures.  Previous microscopy studies\cite{Niu} reveal some films with large domains along the principle axes, some films with many smaller domains along the principle axes, and some films with a distribution of very small domains, many of which are not aligned with the easy axes.  The different domain structures they observe depend sensitively on the film thickness and thermal history during growth. It would not be surprising if the collection of independently grown films in the present study sampled a range of domain types.

The present study concentrates on the paramagnetic tail of the curves.  This is clearly distinguished in type II measurements when an a.c. field near 0.14 Oe is used, but the signal to noise ratio is very poor, so that it is not possible to make meaningful fits of the four independent parameters in eq.(\ref{chiexp}).  We have therefore chosen a field amplitude of 0.56 Oe, where both type I and type II measurements have a much better signal to noise ratio.  This choice requires validation in two ways.  First, it will be necessary to confirm that  Im$\chi(T)$ is not appreciable in the fitting range.  Previous studies of critical transitions in 2D Ising films\cite{Dunlavy,Dunlavy2,Dixon} indicate that if Im$\chi(T)$/Re$\chi(T) \leq$ 0.06 in the paramagnetic region, the linear susceptibility is measured.

 Second, it is necessary to confirm that the shape of the paramagnetic tail is not affected by the rate of change of temperature used.  The effect of the heating rate on the measured susceptibility is illustrated in fig.(\ref{fig1}c).  The particular concern is the possible deformation of the shape of the paramagnetic tail by relaxation from a non-equilibrium state if the heating rate is too large.  To quantify this effect, the location of the point of inflection, $T_{infl}$, of the high temperature tail has been found for measurements at different heating rates on the same film.  This is then used to form the measure $\chi(T_{infl})/\chi(T_{peak})$, or ``how far down the curve does the point of inflection occur".  This gives an indication of the range of data to which it will be possible to fit eq.(\ref{chiexp}).  It can be seen that the heating rate of 0.1 K/s used in this study presents no difficulty.

\section{Results and analysis}

\begin{figure}
\scalebox{.55}{\includegraphics{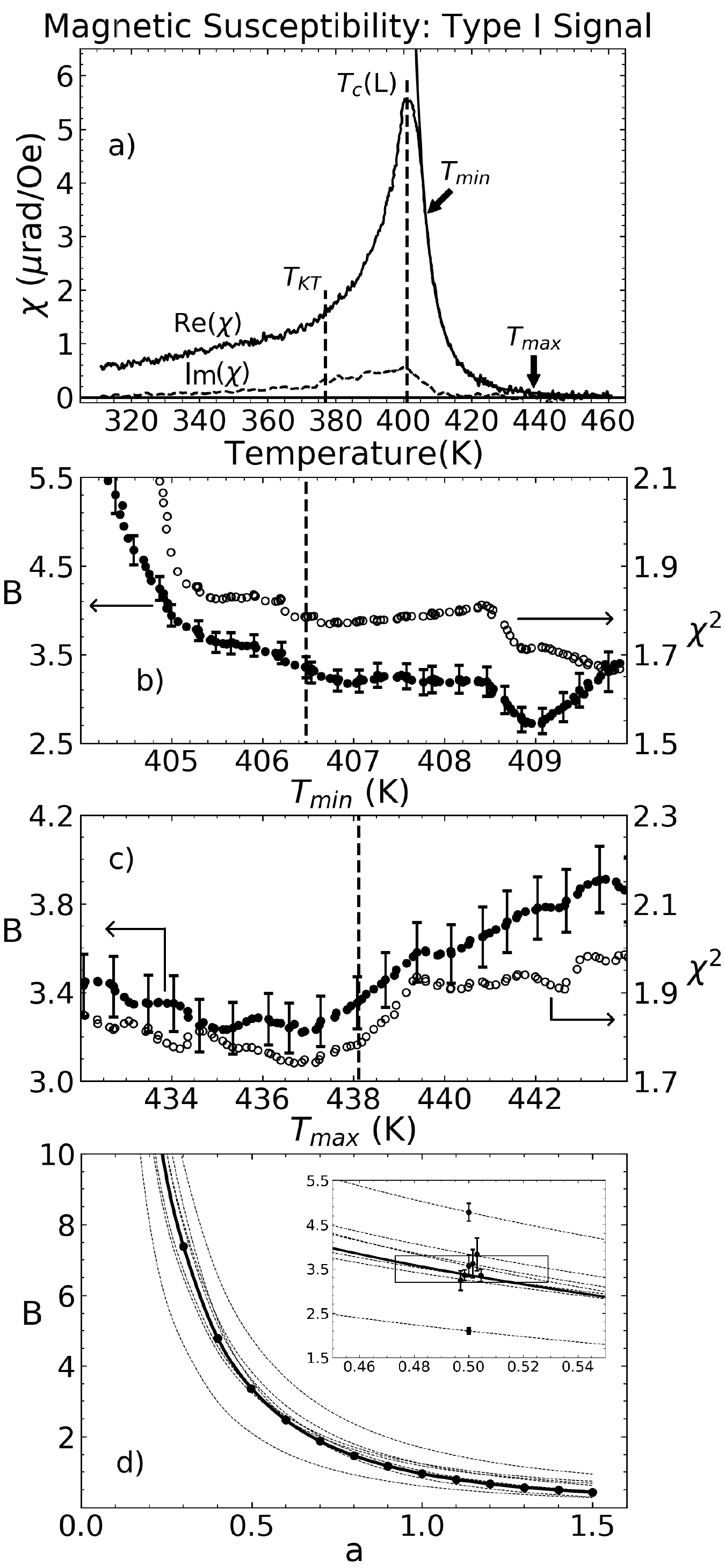}}
\caption{\label{fig2}a) A magnetic susceptibility of type I.  The temperature range of the data that are fit to eq.(\ref{chiexp}) is indicated by $T_{min}$ and $T_{max}$.  The fitted curve is shown with a solid line, and the fitted temperature $T_{KT}$ by a dashed line.  The peak of the susceptibility is indicated by a second dashed line, and labelled $T_C(L)$. b) The fitted value of $B$ (left hand scale, solid symbols) and the $\chi^2$ statistic of the fit (right hand scale, open symbols) as a function of $T_{min}$.  The dashed line shows the value of $T_{min}$ selected for the fit in part a).  c) As in part b), but for the selection of $T_{max}$.  d) The parameter $B$ found in least squares fits to the data in part a), as a function of the parameter $a$, is shown by the points.  The solid line interpolates the points.  A similar fitting procedure yields the interpolated dotted lines for type I measurements from seven additional films.  The inset expands the region near $a=1/2$, and shows the fitted values of $B$, with error bars, for $a=1/2$. }
\end{figure}

Because the type I measurements have a very small imaginary component, they conform most closely to the idealized theoretical model.  Fig.(\ref{fig2}a) shows a representative example of a type I signal from a Fe/W(001) film.   This is a different data set than that shown in fig.(\ref{fig1}).  To determine if this data is described by KT theory, the paramagnetic tail is fit to eq.(\ref{chiexp}).  This requires four parameters to be determined: $\chi_0$, $B$, $T_{KT}$, and the exponent $a$, in addition to a lower and upper limit, $T_{min}$ and $T_{max}$ respectively, to the temperature range where the data is fitted.  

To determine the fitted temperature range, the value $a=1/2$ is chosen. (It was verified that the range does not depend upon this choice.)  A least squares fit to eq.(\ref{chiexp}) is made for $\chi_0$, $B$, and $T_{KT}$ as a function of  $T_{min}$ and $T_{max}$.  Fig.(\ref{fig2}b) and (c) show the fitted value of $B$ and the value of the reduced $\chi^2$ statistic as a function of the bounds.  It can be seen that within the range 406.5 K $< T_{min} < $ 408.7 K, both the fitted value of  $B$ and the $\chi^2$ statistic change very little and are essentially independent of the bounds within the fitted uncertainty.  Close inspection of the data indicates that if $T_{min}$ is moved closer to the susceptibility peak than 406.5 K, it approaches the point of inflection in the data curve where no diverging function will fit well.  If $T_{min}$ is moved further into the tail than 408.7K, it starts to exclude a large proportion of the curve with a high signal-to-noise ratio.  This independence of the fit inside a temperature range that is clearly identified by extrinsic factors is precisely what is expected for a fit that properly represents the data.  $T_{min}$ is therefore chosen on the lower edge of this range, on the principle that maximizing the range will reduce the error in the fitted parameters.  When $T_{min}$ is determined in this way, the average value of Im$\chi(T_{min})$/Re$\chi(T_{min})$ is 0.06$\pm$0.02 for all the type I measurements.  This confirms a linear response in the susceptibility.

Similar criteria are applied to the choice of the upper temperature bound, as shown in fig.(\ref{fig2}c).  In this case, moving $T_{max}>$ 438 K starts to include very noisy data in the fit.  There are 710 data points in the selected fitting region for this data set.

Having determined the fitting range, least squares fits are made for $\chi_0$, $B$, and $T_{KT}$ for a selection of values of $a$ from 0.10 to 1.50.  The values of $B$ are plotted in fig.(\ref{fig2}d) as a function of $a$, with the solid line interpolating the values as a guide to the eye for the data set in fig.(\ref{fig2}a).  The value of the reduced statistical $\chi^2$ for the best fit is essentially independent of the value of $a$, and provides no statistical basis for determining $a$.  However, because KT theory gives independent predictions of $a$ and $B$, both must be met simultaneously.   The fact that the interpolated curve passes through the theoretically predicted range $a=1/2$, $3.2<B<3.8$, shows that the data is consistent with KT theory.  The fitted line is shown in fig.(\ref{fig2}a);   it is mostly obscured by the data itself.  The fitted $T_{KT}$ is indicated by a vertical dashed line.

Fig.(\ref{fig2}d) includes the results of fitting $B$ as a function of $a$ for 7 further measurements of type I signals as interpolated dotted lines.   The inset gives more detail near $a=1/2$.  Six of the eight curves pass through the small box that represents the theoretical range of $B$, and give an average value $B=3.49$ with a standard deviation of 0.22 when $a$=1/2.  Two curves are not in quantitative agreement with KT theory.  Given that the fitted values of $B$ lie many standard deviations from the other six curves, this is not likely to be a statistical variation.  However, as we cannot identify a systematic experimental explanation for these outliers, we have no reason to exclude them from the figure.  Including these in the average yields $B=3.48\pm0.74$.  The width of the box along the horizontal axis establishes a conservative estimate of the uncertainty in $a$.  The top left and bottom right corners of the box are situated so that the average values of $B(a)$ for the six measurements pass through them.  This determines $a=0.50\pm0.03$.

Because the type I measurements have a very small dissipative, imaginary component of susceptibility, the position of the peak of the curve should fairly represent the finite-size transition temperature $T_C(L)$.  A striking feature in fig.(\ref{fig2}a) is that the fitted value of $T_{KT}$ is more than 20 K below the peak maximum.  This is precisely the behaviour expected for a finite-size KT transition, and is expressed quantitatively by eq.(\ref{TC}).  The average value of $T_C(L)/T_{KT}-1$ for the eight data sets is $0.065\pm0.016$, which gives an estimate of the finite size $L \approx 1 \mu$m.  A physical parameter of this order of magnitude is the dimension of the magnetic domains.  These observations give experimental support to the idea that this 2DXY system has very significant finite-size effects even in mesoscopic samples, and that Fe/W(001) ferromagnetic films support a finite-size KT transition.

Finally, these results yield a value for $\eta$ in the temperature range 20-30 K above $T_{KT}$ where eq.(\ref{chiexp}) is fit to the data.  Using the best value of $B$ (without outliers) yields $\eta = 0.12\pm0.09$.  There is a large uncertainty, but it seems that $\eta$ is reduced from the value of 0.25 expected at $T_{KT}$.

\begin{figure}
\scalebox{.4}{\includegraphics{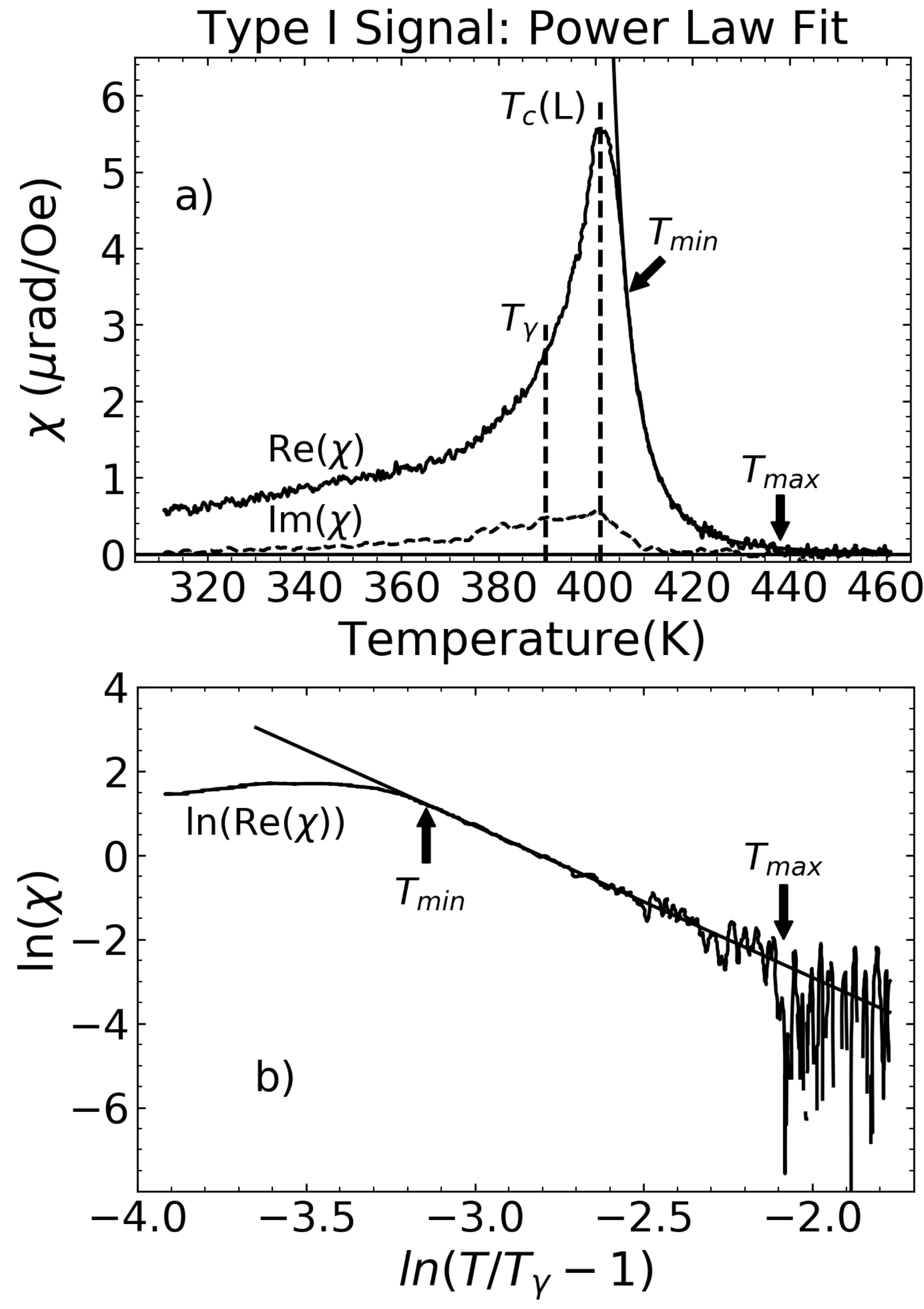}}
\caption{\label{fig3} The data in fig.(\ref{fig2}a) are fit to the power law in eq.(\ref{power}) appropriate for a second order transition.  a) Re$\chi(T)$ is fit using three independent parameters: $\chi_0$, $\gamma$, and the second order transition temperature $T_{\gamma}$.  The solid line illustrates the high quality fit, with a statistical $\chi^2$ that is essentially indistinguishable from the fit to KT theory.   Dashed lines indicate $T_{\gamma}$ and the peak maximum at $T_C(L)$.  b) The same fit is shown on logarithmic scales, where the slope of the fitted line gives the exponent $\gamma = 3.61 \pm 0.08$.  }
\end{figure}

The eight type I data sets have also been fitted to eq.(\ref{power}), as would be appropriate for a second order critical transition.  In this case, there are only three free parameters, $\chi_0$, $\gamma$, and the second-order transition temperature $T_{\gamma}$.  The designation $T_C(L)$ is retained for the temperature at which the data has a maximum.    Fig.(\ref{fig3}) illustrates the power law fit for the same data as in fig.(\ref{fig2}a).  The statistical $\chi^2$ for this fit is indistinguishable from that for eq.(\ref{chiexp}), and the solid line in part a) is again mostly obscured by the data itself.  Fig.(\ref{fig3}b) shows the fit on a log-log scale.  However, the fitted parameters are unphysical in two respects.  First, the value of $\gamma=3.61 \pm 0.08$ is much larger than that of the 2D Ising model or 4-state Potts model, and does not correspond to any known universality class.  Second, the fitted second-order transition temperature $T_{\gamma}=389.7 \pm 0.5$ K is 12 K below the peak temperature, and appears almost halfway down the low-temperature side of the susceptibility peak.\footnote{If $T_{\gamma}$ is not fit, but set equal to $T_C(L)$, then the statistical $\chi^2$ of the fit is worse by a factor of 3.5 and is rejected.} This is very different from the second order 2D Ising phase transition of Fe/W(110), where the fitted Curie temperature is below, but within a degree K of the susceptibility peak.\cite{Back, Elmers3,Dunlavy,Dunlavy2}  The fit to eq.(\ref{chiexp}) in fig.(\ref{fig3}) is typical.  When the eight type I data sets are fit to a power law, the average fitted parameters are $\gamma = 3.5 \pm 0.8$, and $T_C(L)/T_{\gamma} -1 = 0.027 \pm 0.009$.  

The quantitative agreement between the paramagnetic susceptibility and the finite-size KT transition for type I films motivates a closer analysis of the type II measurements.  Low field measurements of these films, as in fig.(\ref{fig1}b), suggests that they comprise a high temperature response similar to those of type 1, plus a much stronger lower temperature response that is related to the response of magnetic domains.  The first part of this suggestion can be tested by analysing them using the same procedures as used for the type I measurements.

\begin{figure}
\scalebox{.4}{\includegraphics{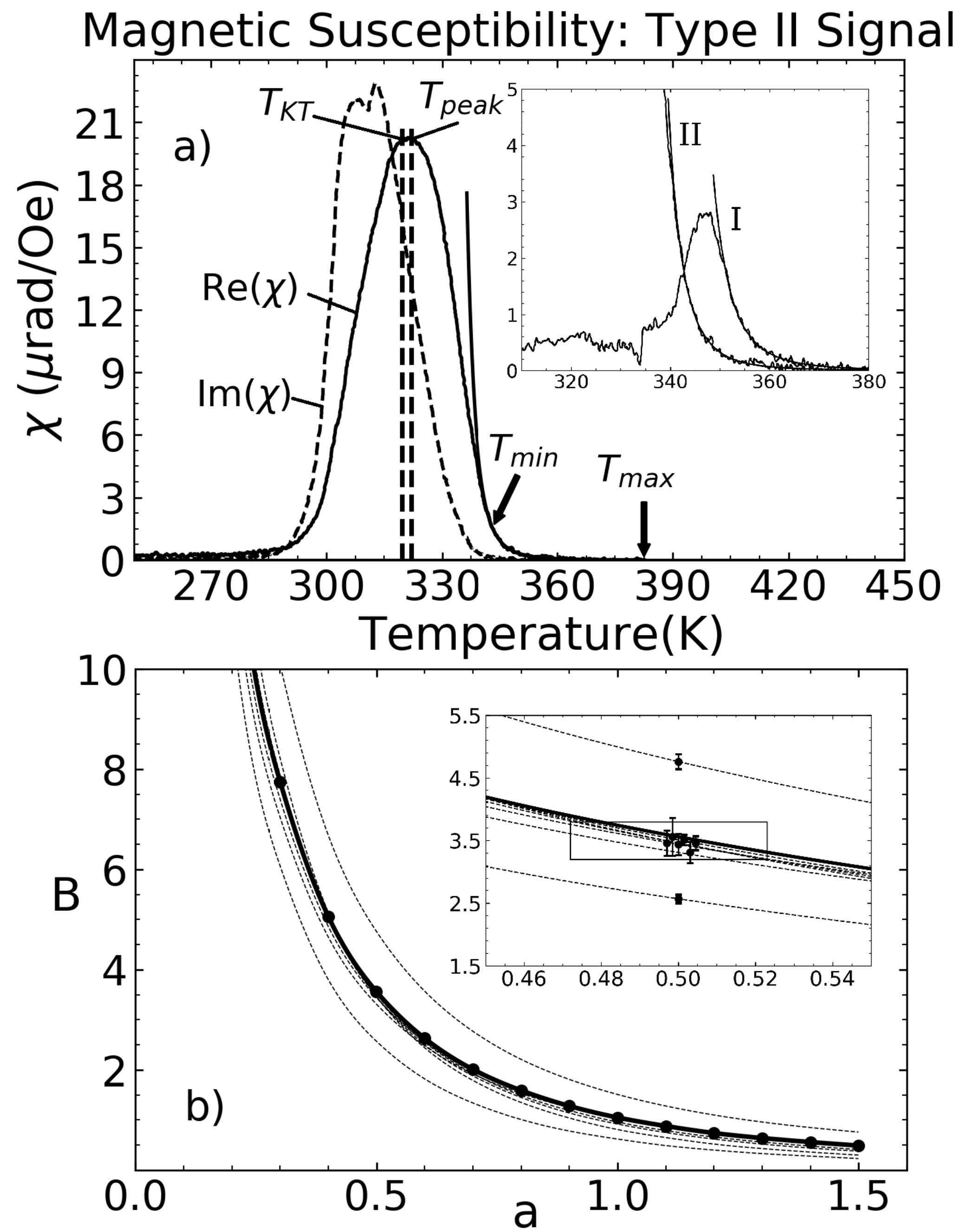}}
\caption{\label{fig4} a) A magnetic susceptibility of type II.  The temperature range used to fit to eq.(\ref{chiexp}) is bounded by $T_{min}$ and $T_{max}$.  The dashed lines indicate the fitted value of $T_{KT}$ and the maximum of the curve, $T_{peak}$.  The solid line shows the fit to the paramagnetic tail of the data.   The inset compares the fit of the type II measurement in the main panel to a type I measurement from the same film, but along an easy axes at right angles to the first.  The two fits have comparable ranges in temperature and the magnitude of the susceptibility.  b) The fitted values of $B$ as a function of chosen values of $a$ for the data in part a) are plotted as points with an interpolating solid line.  A similar fitting procedure yields the interpolated dotted lines for  type II measurements from seven additional films.  The inset expands the region near $a=1/2$, and shows the fitted values of $B$, with error bars, for $a=1/2$.}
\end{figure}
Figure (\ref{fig4}a) shows a susceptibility curve of type II, where there is a large dissipative component Im$\chi(T)$, and Re$\chi(T)$ is about an order of magnitude larger than in measurements of type I.  Using the same criteria for the fitting range as in fig.(\ref{fig1}) gives fitted parameters independent of the range only for the extreme paramagnetic tail.  This is because the point of inflection of Re$\chi(T)$ moves down the curve until the fitted region contains only a very small dissipative component.  For the eight type II measurements that are analyzed, the average value of Im$\chi(T_{min})$/Re$\chi(T_{min})$=0.07$\pm$0.02, again confirming that the susceptibility is linear within the fitted range.  The inset to fig.(\ref{fig4}a) shows the fitted regions of two curves measured from the same film; one is the type II signal in the main panel, and the other is a type I signal measured from the same film, but for the magnetization component along the easy axis at right angles to the first.  The two fits actually cover comparable ranges in temperature and in the magnitude of the susceptibility, and provide equivalent parameters (within uncertainty), other than the fact that there is a shift of about 5 K in the absolute position of the curves.  This supports the speculation that type II measurements are equivalent to type I measurements with the addition of a strong dissipation mechanism below $T_C(L)$.  The dashed lines in the main panel indicate the fitted value of $T_{KT}$ and the maximum of the curve $T_{peak}$.  The maximum is not labelled $T_C(L)$ because the large dissipation in this temperature range is expected to alter the peak shape and shift the location of the maximum from the finite-size transition temperature $T_C(L)$.

Figure (\ref{fig4}b) is a summary plot of the fitted values of $B$ as a function of $a$ for 8 susceptibility measurements of type II.  The solid line corresponds to the data in fig.(\ref{fig4}a), and the dotted lines to the others.  Once again, six out of eight of the curves interpolating the fitted values of $B$ as a function of $a$ pass through the small box representing the theoretical range of $B$ for $a=1/2$.  For these measurements $B=3.46\pm 0.08$.  Including the two curves that do not go through the box gives $B=3.5 \pm 0.6$. The range in $a$ for the type II measurements, established again by the mean of the curves passing through the theoretical limits of B, is $a=0.50 \pm0.03$.  Fitting the type II measurements instead to a power law with three free parameters $\chi_0$, $\gamma$ and $T_{\gamma}$, gives an average value of $\gamma_{eff} = 3.9\pm0.5$, consistent with the type 1 measurements, but not consistent with any known universality class of second order transitions.  Because $T_C(L)$ cannot be determined in the presence of strong dissipation,  fitted values of $T_{KT}$ and $T_{\gamma}$ for the type II data cannot be properly normalized across the eight data sets.  As a result, a meaningful average value of these temperatures cannot be calculated.

These results show that there is no significant difference between the type I and type II curves in the paramagnetic region.  As a result, they can be combined to give a best value of $B=3.48 \pm 0.16$ (excluding 4 outliers), and $a=0.50\pm0.03$.  In summary, the paramagnetic tail of the magnetic susceptibility of Fe/W(001) films is described to a high degree of accuracy by the theory for a finite-size KT transition to a gas of vortices and antivortices.

\section{Conclusions}
Magnetic susceptibility measurements of 3-4 ML Fe/W(001) films have demonstrated that a finite-size KT transition occurs in this 2DXY ferromagnetic system, despite the anisotropies that are present in real samples.    The measurements on many independently grown films follow the predicted exponential form of the paramagnetic susceptibility, with quantitative constants $a=0.50\pm0.03$ and $B=3.46\pm0.16$ in agreement with KT theory.  Stated more exactly, the measurements yield a tight correlation between the parameters $a$ and $B$ in eq.(\ref{chiexp}), but there is no purely statistical basis for choosing any point on the correlation curve.  However, KT theory predicts values of $a$ and $B$ independently, such that two tests of the applicability of the theory may be made.  Given the predicted value of $a$, $B$ is determined independently to be in agreement with KT theory with a small range of uncertainty.  Given the predicted range of $B$, $a$ is  determined independently to be in agreement with KT theory with a small range of uncertainty.  We know of no theoretical description that would support different values of $a$ and $B$.  Fitting the paramagnetic susceptibility to a power law, as predicted for a second-order critical transition, leads to unphysical results.

For the subset of measurements where the dissipation is small (type I), the peak of the susceptibility can be identified as the finite-size transition temperature $T_C(L)$.  The fitted value of $T_{KT}$ is substantially below $T_C(L)$, as predicted for a finite-size KT system.  These samples have  an average value of $T_C(L)/T_{KT}-1 =0.065\pm0.016$.  This gives an estimate of the finite size affecting the transition as $L \approx 1 \mu$m.  This is the order of magnitude of the dimension of magnetic domains, and is consistent with the idea that a mesoscopic limitation of the vortex anti-vortex gas leads to the finite-size transition.

There are a very few previous reports of measurements of the magnetic susceptibility of ultrathin films grown on (001) or (111) cubic surfaces to which these results can be compared.\cite{Taroni}  Susceptibility measurements of Co/Cu(001) films have a substantially wider peak than, for instance, those of Fe/W(110).\cite{Back2}  This might be consistent with a KT transition, but the data is not suitable for a quantitative analysis.   Measurements of Fe/GaAs(001) indicate a very narrow susceptibility\cite{Bensch} with a width of about 1K when an a.c. field of 0.04 Oe is used.  This is not consistent with a finite-size KT transition. In perpendicularly magnetized films, such as Ni/Cu(001),\cite{Poulopoulos} dipole interactions play an important role and change the characteristics of the transition to paramagnetism.  This leaves only studies of Fe/W(001) films.\cite{Elmers3,Elmers1}  Investigations of the effective critical exponent $\beta_{eff}$ of the magnetization between $T_{KT}$ and $T_C(L)$ report a value of $0.22 \pm 0.03$, in agreement with the prediction for a finite-size KT transition.\cite{Bramwell1}  A single measurement of the susceptibility of Fe/W(001) has been published.\cite{Elmers1}  The shape of the peak is very similar to the type I measurements presented here, but  the analysis performed on this data using KT theory is problematic.  The comparison to a second-order transition, yields a value of $\gamma \approx 5$.  As in the present study, this value is unphysically large.

In summary, the present study has established that the paramagnetic susceptibility of Fe/W(001) thin films is described quantitatively by finite-size KT theory which assumes a vortex-antivortex paramagnetic state.  Coupled with previous results that show that this system exhibits finite-size KT scaling of the magnetization between $T_{KT}$ and $T_C(L)$, this is strong experimental confirmation that the KT transition exists in imperfect ultrathin  ferromagnetic films.  The ability to experimentally realize and study the KT transition in a 2DXY system using a simple and accessible ultrathin ferromagnetic film sample will allow the detailed investigation of many interesting questions concerning the topological states and transitions in 2D magnetism.  Among these is the process by which the system moves from discrete four-fold anisotropy to a state of emergent isotropy. The current study hints that magnetic susceptibility measurements will be a sensitive method to explore the crossover from a regime dominated by domain wall excitations to one dominated by the breaking of vortex-antivortex pairs, and that the details of the crossover are related to the underlying distinction between type I and type II susceptibility signals.  Other questions that can be addressed include the relaxation dynamics near and above $T_C(L)$, and the possibility of asymmetry between pair breaking and pair binding dynamics that is a common feature of KT systems.

\begin{acknowledgments}
Financial support for this work  was provided by the Natural Sciences and Engineering Research Council of Canada (NSERC) through the Discovery program.   J. A. acknowledges an Ontario Graduate Scholarship from the province of Ontario. B. N. acknowledges an Undergraduate Student Research Award from NSERC.
\end{acknowledgments}

\bibliography{KT_transition}

\end{document}